\documentclass[11pt,twoside]{article}
\usepackage{iap2000,psfig}%
\newcommand{\simgt}{\lower.5ex\hbox{$\; \buildrel > \over \sim \;$}}
\newcommand{\simlt}{\lower.5ex\hbox{$\; \buildrel < \over \sim \;$}}
\newcommand{\A}{{\rm\scriptscriptstyle A}}
\newcommand{\B}{{\rm\scriptscriptstyle B}}
\newcommand{\gas}{{\rm\scriptscriptstyle gas}}
\newcommand{\bol}{{\rm\scriptscriptstyle bol}}
\newcommand{\band}{{\rm\scriptscriptstyle band}}
\newcommand{\ns}{Log$N$--Log$S$~}
\newcommand{\unit}{erg cm$^{-2}$ s$^{-1}$}
\begin{document}

\title{THE UNIVERSE TRACED BY CLUSTERS
\\[16pt]
%\mbox{%\includegraphics[]{author1.eps}}
}

\author{Yasushi Suto}

\address{Department of  Physics 
and Research Center for the Early Universe\\
The University of Tokyo, Tokyo 113-0033, Japan}

\maketitle

\abstracts{ I critically review a methodology of using clusters of
  galaxies as cosmological probes.  The understanding of the
  abundances and spatial correlations of dark matter halos has been
  significantly advanced especially for a last few years. Nevertheless
  such dark matter halos are not necessarily identical to clusters of
  galaxies in the universe. This is a quite obvious fact, but it seems
  that the resulting systematic errors, on the cosmological parameter
  estimates for instance, are often neglected in the literature.}

\section{Why clusters ?}

This is probably an easy question to answer. In fact, there are
several reasons why clusters of galaxies are regarded as useful probes
of cosmology, including (i) since dynamical time-scale of clusters is
comparable to the age of the universe, they should retain the
cosmological initial condition fairly faithfully, (ii) clusters can be
observed in various bands including optical, X-ray, radio, mm and
submm bands, and in fact several on-going projects aim to make
extensive surveys and detailed imaging/spectroscopic observations of
clusters, (iii) to the first order, clusters are well approximated as
a system of dark matter, gas and galaxies, and thus theoretically
well-defined and relatively well-understood, at least compared with
galaxies themselves, and (iv) on average one can observe a higher-$z$
universe with clusters than with galaxies.  It is established that
X-ray observations are particularly suited for the study of clusters
since the X-ray emissivity is proportional to $n_e^2$ and thus less
sensitive to the projection contamination which has been known to be a
serious problem in their identifications with the optical data. Also
the recent progress of interferometric mapping technique of clusters
via the Sunyaev-Zel'dovich effect enables one to observe the
high-redshift clusters without suffering from the cosmological dimming
$\propto (1+z)^{-4}$.

For those reasons, there are many, perhaps already {\it too many},
papers which discuss cosmological implications of clusters of galaxies
from various statistical methods. Table \ref{tab:list} lists
representative topics of cosmological studies of clusters, with some
possible bias to my personal interest, and thus in particular the
reference list is far from complete.

In what follows, I discuss cosmological implications of the Sunyaev --
Zel'dovich effect (\S \ref{sec:sz}) and cluster abundances (\S
\ref{sec:abundance}), paying attention to the limitations and possible
systematics of current theoretical modeling of galaxy clusters.

%%%%%%%%%%%%%%%%%%%%%%%%%%%%%%%%%%%%%%%%%%%%%%%%%%%%%%%%%%%%%%%%%%%%%%%
\begin{table}[h]
\caption{An incomplete list of ``cosmology with clusters'' \label{tab:list}}
\vspace{0.4cm}
\begin{center}
\begin{tabular}{|l|c|c|}\hline
topic & quantities & references \\ \hline
distance indicator & $H_0$, $\Omega_0$, $\lambda_0$& ~1 -- ~5 \\ \hline
peculiar velocity field & $v_{\rm pec}$& ~6 -- ~9 \\ \hline
mass, temperature and luminosity functions &  $\sigma_8$, $\Omega_0$
& 10 -- 20 \\ \hline
spatial clustering and its evolution &$\xi(r,z)$,  $b(r,z)$ 
& 21 -- 29 \\ \hline
baryon fraction and dark matter &$\Omega_b$,  $\Omega_0$ 
& 30 -- 33\\ \hline
CMB anisotropy through the SZ effect &  $\delta T/T$
& 34 -- 36 \\ \hline
universal density profile/nonlinear clustering  &  $\rho(r)$, $P(k)$ 
& 37 -- 45 \\ \hline
\end{tabular}
\end{center}
\end{table}
%%%%%%%%%%%%%%%%%%%%%%%%%%%%%%%%%%%%%%%%%%%%%%%%%%%%%%%%%%%%%%%%%%%%%%%

\section{Distance and peculiar velocity from 
  the Sunyaev-Zel'dovich effect \label{sec:sz}}

\begin{figure}[h]
\begin{center}
 \leavevmode\psfig{file=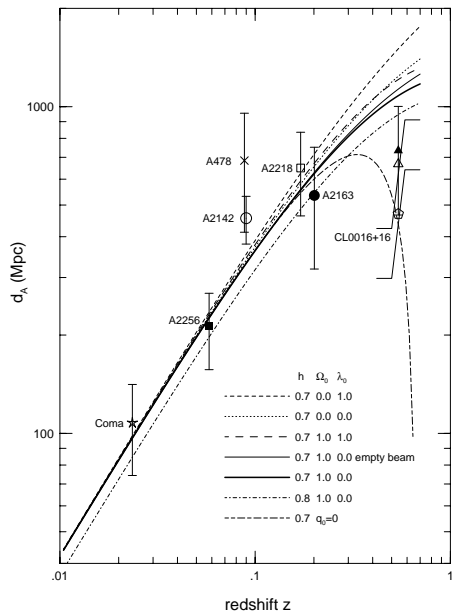,height=8cm}
\end{center}
\caption{The angular diameter distance $d_{\rm A}$ 
  estimated from the SZ effect as a function of $z$ for seven clusters
  (symbols).  Lines indicate theoretical curves for several sets of
  cosmological parameters ($h$, $\Omega_0$, $\lambda_0$). From
  reference 4.
\label{fig:daz}
}
\end{figure}

There have been many attempts to determine the Hubble constant $H_0$
($\equiv 100h\;{\rm km \, s^{-1}\, Mpc^{-1}}$) from the
Sunyaev-Zel'dovich (SZ) temperature decrement and the X-ray
measurements.\cite{SZ72,Birkinshaw,R95} In fact, the observation of
the SZ effect determines the angular diameter distance $d_{\rm A} (z)$
to the redshift $z$ of each cluster\cite{SW78,Kobayashi}.  For $z \ll
1$, $d_{\rm A}(z)$ is basically given only in terms of $H_0$ ($\sim
cz/H_0$, with $c$ being the light velocity).  If $z \simgt 0.1$,
however, the density parameter $\Omega_0$, the dimensionless
cosmological constant $\lambda_0$, and possibly the degree of the
inhomogeneities in the light path make significant contribution to the
value of $d_{\rm A}(z)$ as well. As far as I understand, Figure
\ref{fig:daz} is the first Hubble diagram from the SZ effect published
in the literature.\cite{Kobayashi} Note that the data for CL0016+16
plotted in this figure have been corrected later. Figure \ref{fig:daz}
clearly indicates the potential difficulty of determining the
cosmological parameters from the SZ effect of an individual cluster.
Of course this is not surprising because the estimate of the angular
diameter distance from the SZ effect crucially rests on several
idealistic assumptions including the spherical symmetry, isothermal
temperature distribution, and no clumpiness of the gas. It is quite
ironical that one can immediately point out the possible systematics
by abandoning one of the assumptions because the SZ effect is based on
the well-defined physics rather than on the unjustifiable, and thus
unfalsefiable, empirical result which is the case for the other
distance indicators.

%%%%%%%%%%%%%%%%%%%%%%%%%%%%%%%%%%%%%%%%%%%%%%%%%%%%%%%%%%%%%%%%%%%
%%%%%%%%%%%%%%%%%%%%%%%%%%%%%%%%%%%%%%%%%%%%%%%%%%%%%%%%%%%%%%%%%%%

\begin{figure}[ht]
\begin{center}
    \leavevmode\psfig{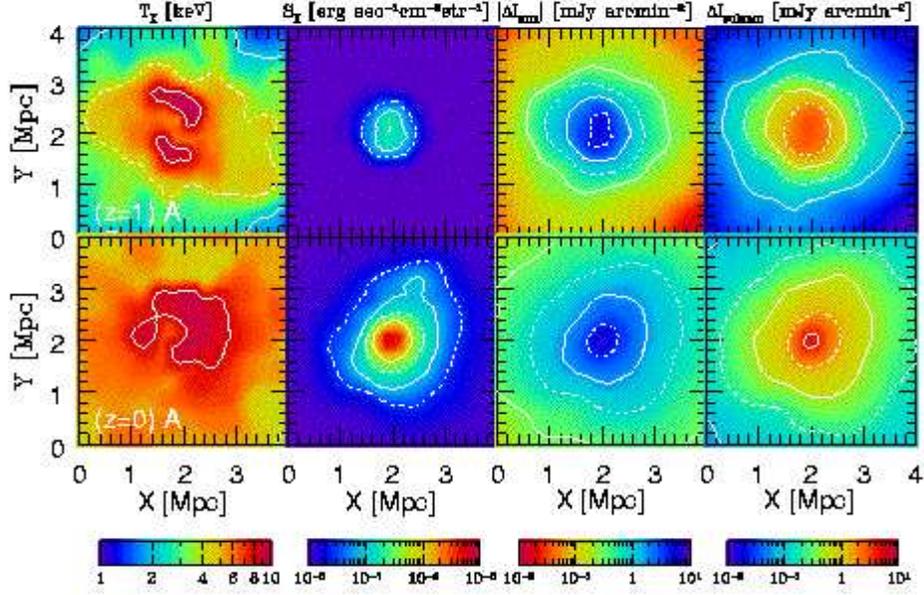}
\end{center}
\caption{Projected views of a simulated clusters 
  at $z=1$ and $z\approx 0$. A box of (4Mpc)$^3$ (in physical lengths)
  located at the center of each cluster is extracted.  The X-ray
  emission--weighted temperature ($T_{\rm X}$), X-ray surface
  brightness ($S_{\rm X}$), and the SZ surface brightness at mm and
  submm bands ($|\Delta I_{\rm mm}|$ and $\Delta I_{\rm submm}$) are
  plotted on the projected X-Y plane by integrating over the
  line-of-sight direction (Z). The X and Y coordinates are in the
  physical lengths at the corresponding redshift, and related to the
  angular coordinate $\theta$ from the cluster center as $\theta
  d_{\rm A}(z)$. At $z\approx 0$, $d_{\rm A}(z)$ can be replaced by
  the real distance to the cluster from the observer.  From reference
  9.
\label{fig:yisplate}
}
\end{figure}

\begin{figure}[ht]
\begin{center}
    \leavevmode\psfig{file=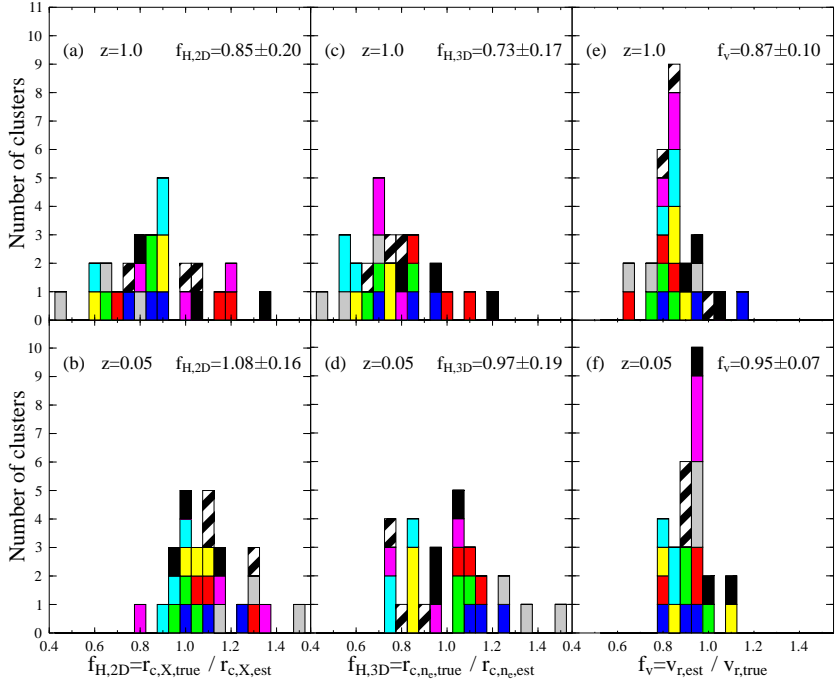,angle=-90,width=11cm}
\end{center}
\caption{Distribution of $f_{H,{\rm 2D}}$ ({\it left}), 
  $f_{H,{\rm 3D}}$ ({\it middle}), and $f_v$ ({\it right}) for all of
  the simulated clusters (nine in total) at $z=0.05$ and $z=1.0$
  viewed from three different line-of-sight directions. Different
  patterns of the histogram correspond to different clusters. The mean
  and 1$\sigma$ statistical errors are quoted in each panel.  From
  reference 9.
\label{fig:allhist}}
\end{figure}

Therefore it is essential to accumulate a statistically meaningful
number of the SZ clusters. For this purpose in mind, we performed
Smoothed Particle Hydrodynamical (SPH) simulations of clusters of
galaxies in Cold Dark Matter (CDM) models.\cite{YIS98} Figure
\ref{fig:yisplate} presents an example of a simulated cluster observed
in different bands. This illustrates how an individual cluster
exhibits the departure from the above idealistic assumptions. Figure
\ref{fig:allhist} summarizes the distribution of the estimated Hubble
constant from the projected X-ray surface brightness profile ({\it
  left}) and the 3D density profile ({\it middle}) and of the
estimated peculiar velocity of clusters ({\it right}).\cite{YIS98} The
result suggests that the intrinsic scatter reflecting the different
cluster gas state is fairly large, especially at high redshifts, and
that at least a few tens of clusters should be observed to determine
the parameters within 10 percent accuracy. Of course this conclusion
is heavily dependent on the extent to which the simulated clusters
faithfully reproduce the statistical properties of the observed
cluster samples in the universe. To properly answer this question, we
should wait for an extensive survey of clusters with {\it high-angular
  resolution}.  Incidentally, our recent SZ image ($\sigma_{\rm FWHM}
=13''$ at 150~GHz) of the most luminous X-ray cluster RX~J1347--1145
($z=0.45$) indeed revealed complex morphological structures of the
cluster region, \cite{komatsu99,komatsu01} and therefore the departure
from the the spherical isothermal $\beta$-model for the clusters may
be more significant than currently thought.

%%%%%%%%%%%%%%%%%%%%%%%%%%%%%%%%%%%%%%%%%%%%%%%%%%%%%%%%%%%%%%%%%%%%
%%%%%%%%%%%%%%%%%%%%%%%%%%%%%%%%%%%%%%%%%%%%%%%%%%%%%%%%%%%%%%%%%%%

%%%%%%%%%%%%%%%%%%%%%%%%%%%%%%%%%%%%%%%%%%%%%%%%%%%%%%%%%%%%%%%%%%%%
%%%%%%%%%%%%%%%%%%%%%%%%%%%%%%%%%%%%%%%%%%%%%%%%%%%%%%%%%%%%%%%%%%%%

%%%%%%%%%%%%%%%%%%%%%%%%%%%%%%%%%%%%%%%%%%%%%%%%%%%%%%%%%%%%%%%%%%%%%%
\begin{figure}[ht]
  \begin{center}
    \leavevmode\psfig{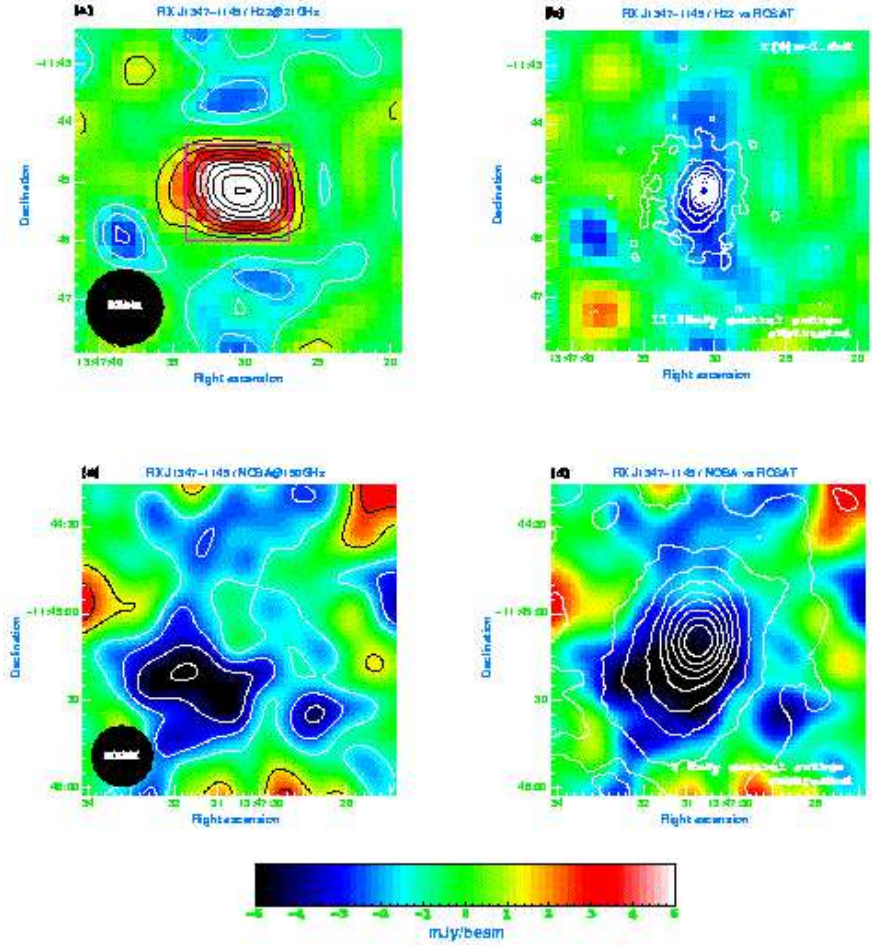}
\end{center}
\caption{
  The SZ maps of RX~J1347--1145.  (a) the 21~GHz map: $6'\times 6'$
  ($2.4~h_{50}^{-1}~{\rm Mpc}\times 2.4~h_{50}^{-1}~{\rm Mpc}$).  (b)
  the 21~GHz map after subtracting the central point source. (c) the
  150~GHz map: $1'9\times 1'9$ ($0.75~h_{50}^{-1}~{\rm Mpc}\times
  0.75~h_{50}^{-1}~{\rm Mpc}$). (d) the 150~GHz map after subtracting
  the central point source (assuming the flux of 3.8~mJy) overlaid
  with the X-ray contours. From reference 48.
\label{fig:szmap}
}
\end{figure}
%%%%%%%%%%%%%%%%%%%%%%%%%%%%%%%%%%%%%%%%%%%%%%%%%%%%%%%%%%%%%%%%%%%%%%

\section{Cluster abundance \label{sec:abundance}}

Cosmological implications of cluster abundance are discussed using a
variety of statistics including X-ray temperature
function,\cite{HA91,WEF,VL,ECF,KS96} mass function,\cite{BC,UIS}
velocity function\cite{shimasaku,usss} and X-ray luminosity function and
log $N$ - log S relation.\cite{EH91,B91,Barbosa,KS97,oukbir,Borgani99b}.
In what follows, we specifically consider the theoretical model
\cite{KS96,KS97,KSS} for X-ray log $N$ - log S relation in order to
illustrate the systematic effect in cosmological conclusions derived
from cluster abundance.

The number of clusters observed per unit solid angle with X-ray flux
greater than $S$ is predicted from
%%%%%%%%%%%%%%%%%%%%%%%%%%%%%%%%%%%%%%%%%%%%%%%%%%%%%%%%%%%%%%
\begin{eqnarray}
  N(>S)= \int_{0}^{\infty}dz ~d_\A^2(z) \, c
  \left|{\frac{dt}{dz}}\right| \int_{S}^\infty dS ~ (1+z)^3 n_M(M,z)
  \frac{dM}{dT_{\gas}}\frac{dT_{\gas}}{dL_\band} \frac{dL_\band}{dS},
\label{eq:logns}
\end{eqnarray}
%%%%%%%%%%%%%%%%%%%%%%%%%%%%%%%%%%%%%%%%%%%%%%%%%%%%%%%%%%%%%%
where $c$ is the speed of light, $t$ is the cosmic time, $d_\A$ is the
angular diameter distance, $T_{\gas}$ and $L_{\band}$ are respectively
the gas temperature and the band-limited absolute luminosity of
clusters, and $n_M(M,z)dM$ is the comoving number density of
virialized clusters of mass $M \sim M+dM$ at redshift $z$ (we use the
Press--Schechter mass function).

Given the observed flux $S$ in an X-ray energy band [$E_a$,$E_b$], the
source luminosity $L_\band$ at $z$ in the corresponding band
[$E_a(1+z)$,$E_b(1+z)$] is written as
%%%%%%%%%%%%%%%%%%%%%%%%%%%%%%%%%%%%%%%%%%%%%%%%%%%%%%%%%%%%%%
\begin{equation}
  L_\band[E_a(1+z),E_b(1+z)] = 4 \pi 
d_{\rm\scriptscriptstyle L}^2(z) S[E_a,E_b],
\label{eq:ls}  
\end{equation}
%%%%%%%%%%%%%%%%%%%%%%%%%%%%%%%%%%%%%%%%%%%%%%%%%%%%%%%%%%%%%%
where $d_{\rm\scriptscriptstyle L} = (1+z)^2 d_\A$ is the luminosity
distance. We adopt the observed $L_\bol - T_\gas$ relation parameterized
by
%%%%%%%%%%%%%%%%%%%%%%%%%%%%%%%%%%%%%%%%%%%%%%%%%%%%%%%%%%%%%%
\begin{equation}
  L_\bol = L_{44} \left( \frac{T_{\gas}}{6{\rm keV}} 
\right)^{\alpha}
  (1+z)^\zeta ~~ 10^{44} h^{-2}{\rm ~ erg~sec}^{-1} .
\label{eq:lt}
\end{equation}
%%%%%%%%%%%%%%%%%%%%%%%%%%%%%%%%%%%%%%%%%%%%%%%%%%%%%%%%%%%%%%
We take $L_{44}=2.9$, $\alpha=3.4$ and $\zeta=0$ as a fiducial set of
parameters.  Then we translate $L_\bol(T_{\gas})$ into the
band-limited luminosity $L_\band[T_{\gas},E_1,E_2]$ taking account of
the thermal bremsstrahlung and the metal line emissions.

Assuming that the intracluster gas is isothermal, its temperature
$T_{\gas}$ is related to the total mass $M$ by
%%%%%%%%%%%%%%%%%%%%%%%%%%%%%%%%%%%%%%%%%%%%%%%%%%%%%%%%%%%%%%%%%%%%%%%%%%
\begin{eqnarray}
  k_\B T_{\gas} &=& \gamma {\mu m_p G M \over 3 r_{\rm vir}(M,z_f)},
  \nonumber \\ &=& 5.2\gamma (1+z_f) \left({\Delta_{\rm vir} \over
      18\pi^2}\right)^{1/3} \left({M \over 10^{15} h^{-1} M_\odot}
  \right)^{2/3} \Omega_0^{1/3} ~{\rm keV}.
\label{eq:tm}
\end{eqnarray}
%%%%%%%%%%%%%%%%%%%%%%%%%%%%%%%%%%%%%%%%%%%%%%%%%%%%%%%%%%%%%%%%%%%%%%%%%%   
where $\mu$ is the mean molecular weight (we adopt $\mu=0.59$), and
$\gamma$ is a fudge factor of order unity.  The virial radius $r_{\rm
  vir}(M,z_f)$ of a cluster of mass $M$ virialized at $z_f$ is
computed from $\Delta_{\rm vir}$, the ratio of the mean cluster
density to the mean density of the universe at that epoch.  
Further we assume that the gas temperature evolves as
%%%%%%%%%%%%%%%%%%%%%%%%%%%%%%%%%%%%%%%%%%%%%%%%%%%%%%%%%%%%%%%%%%%%%%%
\begin{eqnarray} 
  T_{\gas}(M,z_{\rm f},z) = \left(\frac{1+z_{\rm f}}{1+z}\right)^s  
                T_{\gas}(M,z_{\rm f}),
\label{eq:gastemp}
\end{eqnarray} 
%%%%%%%%%%%%%%%%%%%%%%%%%%%%%%%%%%%%%%%%%%%%%%%%%%%%%%%%%%%%%%%%%%%%%%%%
taking account of the quiescent accretion of matter after the formation
epoch $z_{\rm f}$.\cite{KS96}

\begin{figure}[ht]
\begin{center}
  \leavevmode\psfig{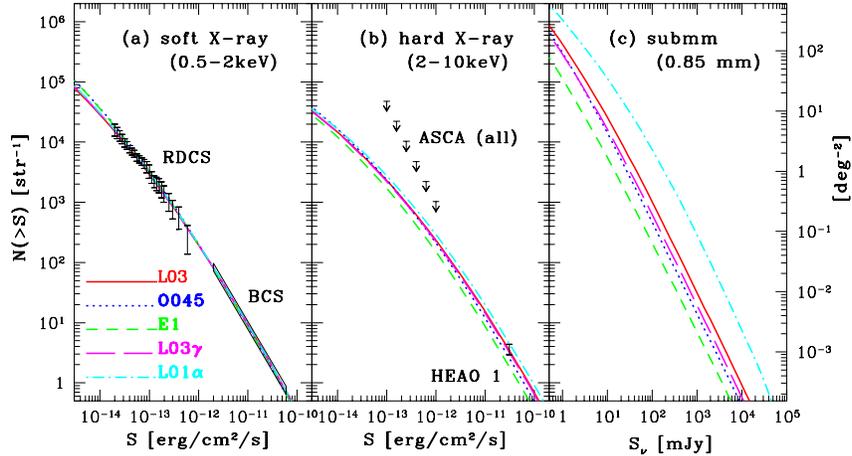}
\end{center}
\caption{The \ns relations of galaxy clusters for CDM models  
  in (a) the soft X-ray (0.5-2.0 keV) band, (b) the hard X-ray (2-10
  keV) band, and (c) the submm (0.85 mm) band. Lines represent the
  models listed in Table 2. From reference 20.
\label{fig:logns3}
}
\end{figure}

The left panel of Figure \ref{fig:logns3} shows the predictions for
the X-ray \ns relations in various CDM models. As is well known now,
many models with appropriate sets of cosmological parameters account
for the observed cluster abundance.  Several examples of such models
are listed in Table \ref {tab:cdmmodel} which exhibit the almost
indistinguishable predictions for the X-ray \ns relations in the ROSAT
band.

%%%%%%%%%%%%%%%%%%%%%%%%%%%%%%%%%%%%%%%%%%%%%%%%%%%%%%%%%%%%%%%%%%%%%%%%%%
\begin{table}[ht]
\caption{CDM models consistent with the {\it ROSAT} X-ray
  Log $N$ -- Log $S$ relation. \label{tab:cdmmodel}}
\begin{center}
\begin{tabular}{|ccccccc|}
\hline
Model & $\Omega_0$ &  $\lambda_0$  
&  $h$ &   $\sigma_8$ & $\alpha$ & $\gamma$\\ \hline 
L03 & 0.3  & 0.7 & 0.7 & 1.04 & 3.4 & 1.2  \\
O045 & 0.45 & 0   & 0.7 & 0.83 & 3.4 & 1.2  \\
E1 & 1.0  & 0   & 0.5 & 0.56 & 3.4 & 1.2 \\
L03$\gamma$ & 0.3  & 0.7 & 0.7 & 0.90 & 3.4 & 1.5 \\
L01$\alpha$ & 0.1  & 0.9 & 0.7 & 1.47 & 2.7 & 1.2 \\
\hline
\end{tabular}
\end{center}
\end{table}
%%%%%%%%%%%%%%%%%%%%%%%%%%%%%%%%%%%%%%%%%%%%%%%%%%%%%%%%%%%%%%%%%%%%

%%%%%%%%%%%%%%%%%%%%%%%%%%%%%%%%%%%%%%%%%%%%%%%%%%%%%%%%%%%%%%%%%%%%%%
%%%%%%%%%%%%%%%%%%%%%%%%%%%%%%%%%%%%%%%%%%%%%%%%%%%%%%%%%%%%%%%%%%%%%%

\begin{figure}[ht]
\begin{center}
  \leavevmode\psfig{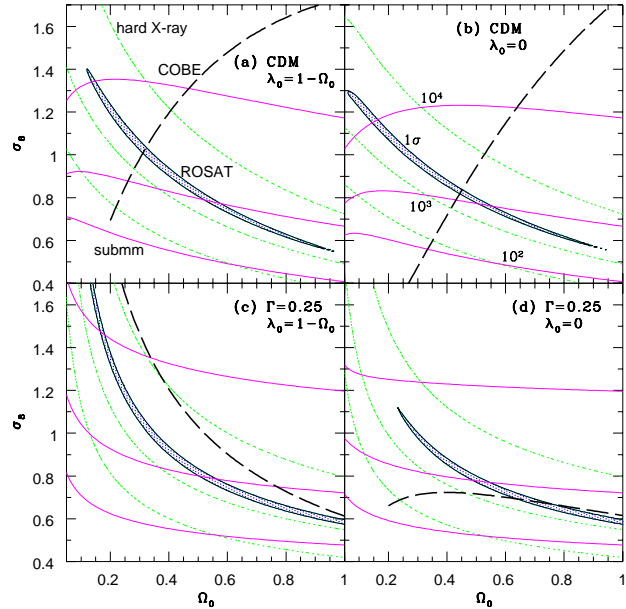}
\end{center}
\caption{Contour maps on the $\Omega_0$-$\sigma_8$ plane in (a) 
  spatially flat ($\lambda_0=1-\Omega_0$) CDM models, (b) open
  ($\lambda_0=0$) CDM models, (c) spatially flat CDM-like models with
  the fixed shape parameter ($\Gamma=0.25$), and (d) open CDM-like
  models with $\Gamma=0.25$.  In all cases, $h=0.7$, $\alpha=3.4$, and
  $\gamma=1.2$ are assumed. Shaded regions represent the 1$\sigma$
  significance contours derived in KS97 from the soft X-ray (0.5-2
  keV) \ns. Dotted and solid lines indicate the contours of the number
  of clusters greater than $S$ per steradian ($10^2$, $10^3$, $10^4$
  from bottom to top) with $S = 10^{-13}$ \unit in the hard X-ray
  (2-10 keV) band and with $S_\nu = 50$mJy in the submm (0.85 mm)
  band, respectively.  Thick dashed lines represent the {\it COBE} 4
  year result. From reference 20.
\label{fig:cont_szac4}
}
\end{figure}

The {\it degeneracy} among those viable cosmological models can be
broken by observing wider (i.e., increasing the statistics), deeper (at
higher redshifts) and/or in multi-wavelength bands.  The middle and
right panels in Figure \ref{fig:logns3} plot the cluster \ns predictions
in hard X-ray (due to the thermal bremsstrahlung), and in submm (due to
the SZ effect). Figure \ref{fig:cont_szac4} illustrates the extent to
which one can break the degeneracy between $\sigma_8$ and $\Omega_0$ in
CDM models ($n=1$, $h=0.7$) using the multi-band observational data.

%%%%%%%%%%%%%%%%%%%%%%%%%%%%%%%%%%%%%%%%%%%%%%%%%%%%%%%%%%%%%%%%%%%%%
%%%%%%%%%%%%%%%%%%%%%%%%%%%%%%%%%%%%%%%%%%%%%%%%%%%%%%%%%%%%%%%%%%%%%

%%%%%%%%%%%%%%%%%%%%%%%%%%%%%%%%%%%%%%%%%%%%%%%%%%%%%%%%%%%%%%%%%%%%
\begin{figure}[ht]
\begin{center}
  \leavevmode\psfig{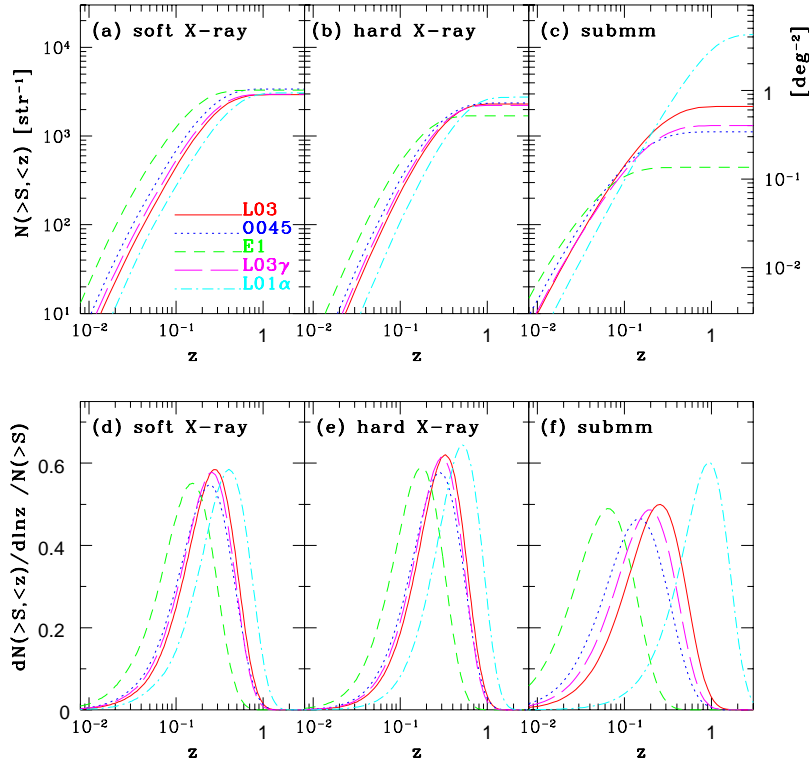} 
\end{center}
\caption{Redshift evolution of the number of galaxy clusters. Upper
  panels show the cumulative number $N(>\!S,<\!z)$ against $z$ in (a)
  the soft X-ray (0.5-2 keV) band with $S=10^{-13}$ \unit, (b) the
  hard X-ray (2-10 keV) band with $S=10^{-13}$ \unit, and (c) the
  submm (0.85 mm) band with $S_\nu=50$ mJy. Lower panels (d)--(f) are
  similar to (a)--(c) except for plotting the differential
  distribution $dN(>\!S,<\!z)/d\ln z$ normalized by $N(>S)$. From
  reference 20.
\label{fig:nsz}
}
\end{figure}
%%%%%%%%%%%%%%%%%%%%%%%%%%%%%%%%%%%%%%%%%%%%%%%%%%%%%%%%%%%%%%%%%%%%%%

Similarly the redshift-distribution of cluster abundances can be a
very powerful discriminator of the different cosmological
models. Figure \ref{fig:nsz} exhibits the redshift evolution of the
number of clusters in different bands. As expected, the evolutionary
behavior strongly depends on the values of $\Omega_0$ and $\sigma_8$;
the fraction of low redshift clusters becomes larger for greater
$\Omega_0$ and smaller $\sigma_8$. This indicates that one may be
able to distinguish among these models merely by determining the
redshifts of clusters up to $z \sim 0.2$. 

Since the above agreement between model predictions and available
observations is so remarkable, the resulting conclusions on the values
of $\Omega_0$ and $\sigma_8$ look robust. Figure \ref{fig:cont_szac4}
explores the possible systematic effects on $\Omega_0$ and $\sigma_8$
by changing the model parameters which describe the cluster gas. For a
realistic range of those parameters, the constraints change around
(10--20) percent level. As a matter of fact, this statement is
somewhat misleading. The crucial assumption underlying the cluster
abundance modeling, I believe, is {\it the one-to-one correspondence
  between the observed galaxy clusters and the dark matter halos}. The
former is defined via the optical luminosity of member galaxies and/or
X-ray luminosity of the cluster gas, while the latter is defined
basically from the spherical collapse model. More specifically, we
have a variety of definitions for {\it clusters}; optically selected
clusters (or the {\it Abell} clusters), X-ray selected clusters, SZ
selected clusters, dark matter halos defined through the spherical
collapse, and halos directly identified from large cosmological
simulations using a variety of selection criteria. They should not be
identical. Of course I agree that assuming the one-to-one
correspondence among those {\it species} is a good approximation. My
point here, however, is that the assumption may easily affect the
derived values of the cosmological parameters more than the systematic
effect presented in the above {\it totally under this assumption}.

I do not think that this has not been considered seriously simply
because the agreement between model predictions and available
observations is {\it satisfactory}. Since current viable cosmological
models are specified by a set of many {\it adjustable} parameters (see
example Table \ref{tab:cdmmodel}), the agreement does not necessarily
justify the underlying assumption. Thus it is dangerous to stop
doubting the unlikely assumption because of the (apparent ?) success.
In this respect, I am always impressed by the marked contrast with the
case of the SZ effect as a distance indicator (\S \ref{sec:sz}), where
the simple model predictions and the observations do not agree
perfectly (with many scatters) and everybody talks of the systematics
carefully enough.

%%%%%%%%%%%%%%%%%%%%%%%%%%%%%%%%%%%%%%%%%%%%%%%%%%%%%%%%%%%%%%%%%%%%%
\begin{figure}[ht]
\begin{center}
 \leavevmode\psfig{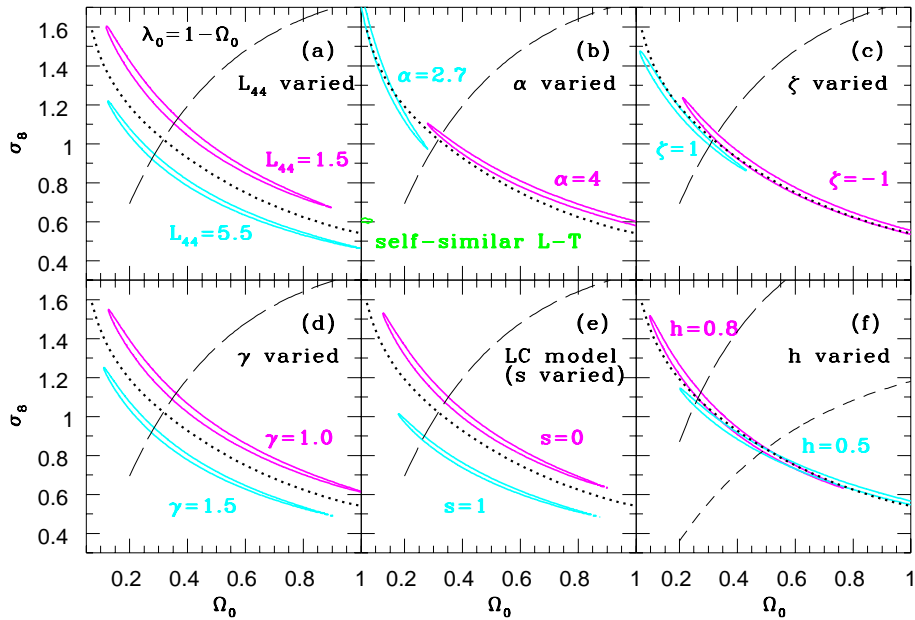}
\end{center}
\caption{Systematic effects on the $\Omega_0 - \sigma_8$ constraints in
  $\lambda_0=1-\Omega_0$ CDM models. The $1\sigma$(68\%) confidence
  contours from the cluster \ns are plotted for different (a) $L_{44}$,
  (b) $\alpha$, (c) $\zeta$, (d) $\gamma$, (e) $s$, and (f) $h$.  Except
  for the parameters varied in each panel, the canonical set of
  parameters ($L_{44}=2.9$, $\alpha=3.4$, $\zeta=0$, $\gamma=1.2$,
  $h=0.7$) and the PS model are used. Dotted and dashed lines represent
  our best-fit for the canonical parameter set and the {\it COBE} 4 year
  results, respectively. From reference 18.
\label{fig:chi2ns6f}
}
\end{figure}
%%%%%%%%%%%%%%%%%%%%%%%%%%%%%%%%%%%%%%%%%%%%%%%%%%%%%%%%%%%%%%%%%%%%%

\section{Summary and conclusions}

Here I presented a brief review of cosmological implications of galaxy
clusters, specifically considering the Sunyaev-Zel'dovich effect and
the cluster abundance in some detail.  I believe that the clusters are
quite important and useful probes of cosmology and in fact they
already proved to be successful in many respects. If one would like to
go further and to extract more stringent constraints on the
cosmological models, however, the one-to-one correspondence between
virialized halos and observed clusters, whatever they mean, should be
critically examined.  This assumption is a reasonable working
hypothesis, but we need more quantitative justification or
modification in order to improve the {\it cosmology with clusters}.

\section*{Acknowledgments}

I thank the organizers, Florence Durret and Daniel Gerbal, for
inviting me to this wonderful meeting.  The present talk is based on
my previous/ongoing work with many collaborators. In particular I
thank Tetsu Kitayama, Eiichiro Komatsu, Shin Sasaki, and Kohji
Yoshikawa.  This research was supported in part by the Grants-in-Aid
by the Ministry of Education, Science, Sports and Culture of Japan
(07CE2002, 12640231).

\end{document}